\documentstyle[aps,pre,multicol,epsf]{revtex}

\begin{document}

\title{Dynamic criticality in driven disordered systems:\\
Role of depinning and driving rate in Barkhausen noise\thanks{Dedicated
to the memory of my father}}

\author{Bosiljka Tadi\'c\thanks{Electronic address: Bosiljka.Tadic@ijs.si}}

\address{Jo\v{z}ef Stefan Institute,
P.O. Box 3000, 1001 Ljubljana, Slovenia }


\maketitle
\begin{abstract}
\newline
We study Barkhausen noise in a diluted two-dimensional Ising model
with the extended domain wall and  weak random fields occurring due to
coarse graining. We report two types of scaling behavior corresponding
to (a) low disorder regime where  a single domain wall slips through
a series of positions when the external field is increased,
 and (b) large disorder regime, which is characterized  with  nucleation of
many domains.  The effects of finite concentration of nonmagnetic ions and
variable driving rate on the scaling exponents is discussed in both regimes.
The universal scaling behavior at low disorder is shown to belong  to a class
of critical dynamic systems, which are  described by a  fixed
point  of the stochastic transport equation with self-consistent disorder
correlations.
\end{abstract}

\begin{multicols}{2}
\section{Introduction}
Barkhausen signal is a physical phenomenon of large practical importance
for noninvasive material characterization technique and for magnetic
recording. It consists of series of irregular pulses (noise) which are related
to reversal of the magnetization by slowly increasing (e.g., from
negative saturation) external field in
ferromagnetic alloys and amorphous ferromagnets.
Recently Barkhausen noise (BN)
has been studied  as an  example of complex behavior
in driven disordered systems far from equilibrium. It has been found
\cite{MC1,TOR,GRB,E4,Torino,BGD,Durin,Durin-stress}
that BN obeys a scale-free behavior  without tuning of a parameter,
a feature which is peculiar to
  the   dynamic critical  systems \cite{Bak}.
In particular, most
often measured are distributions of the area $s$ of Barkhausen pulses,
which  obeys a power-law  up to a cut-off according to
 $D(s) \sim s^{-\tau _s}$, the durations of pulses $t$ with distribution
 $P(t) \sim t^{-\tau _t}$, and the  energy  released in a pulse $n$ with
 $D(n) \sim n^{-\tau _n}$. The
dissipated power spectrum was found to decay with frequency $f$ as
$S(f)\sim f^{-\phi }$. A short
summary of the experimental results taken from the literature
is given in Table\ 1. It should be noted that the
scaling exponents obey (within error bars) the
following scaling relations  $(\tau _s-1)D_s=
(\tau _n-1)D_n=\tau _t-1$, where $D_s$ and $D_n$ are the fractal
dimensions
associated with  the size $s$ and energy $n$ of pulses, respectively.
The measured scaling exponents can be  grouped in (at least)
three distinct universality classes, which we may characterize according
to the exponent $\tau _s$ as 1.33, 1.40, and 1.70,
respectively. The effects of
varying driving rate have  been studied systematically in
certain compounds in Refs.\ \cite{TOR,Durin,Durin-stress,D+BG}.
It has been recognized that increased driving rate
leads to increased occurrence of large pulses  and thus to a continuous
decrease of the scaling exponents.
In addition, a careful analysis of the experimental results suggests that
scaling exponents may depend on sample composition and annealing (compare,
for instance, MC3 and D1 in Table\ 1), and on the exposing of a
magnetostrictive sample to tensile stress, which leads to different
structure of domains \cite{Durin-stress}, as opposed to unstressed sample
\cite{D+BG} (see (DS) and (DB) in Table\ 1).

Proximity of the mean-field universality class
(corresponding to $\tau _s=1.66$, which is  close to measured $\approx
1.7$ in Table\ 1) was understood as being  a consequence of dipolar
forces and demagnetizing
fields in $d=3$ spatial dimensions  \cite{D+BG}.
Recently  occurrence of additional two universality classes
(characterized with $\tau _s=1.54$ and $\tau _s=1.30$)  \cite{TN}
 has been attributed to two-dimensional character of domain-wall motion,
which was observed in  a class of anisotropic soft magnets.
Such systems are for instance ribbons made of Fe-B-Si \cite{MC1,LM} and
 Fe-Co-B compounds \cite{Durin,Durin-stress,D+BG}, in  which domains
 appear to have  stripe structure with
extended domain walls parallel to the magnetization.
The two universality classes have been  related  \cite{TN} to a  single
domain wall motion and to a multidomain structure,
 respectively, in a  2-dimensional prototype model with local random fields.
Models with other types of disorder in $d=2$ dimensions \cite{Vives,BT,BT2}
have considered BN in multidomain structure.

In the present paper  we extend the study of Ref.\ \cite{TN} to  consider
Barkhausen noise in two regimes corresponding to a single domain  and
multi-domain structure, respectively, in a {\it diluted} Ising model
in $d=2$  with varying concentration of
nonmagnetic ions $c$ and fixed (weak) random fields.
We use infinitely slow driving in order to study effects of the
varying concentration $c$ on the
statistics of BN avalanches. We also  discuss the effects of finite
driving rates in both regimes (see also \cite{BT}).
Finally, we show that the critical behavior in the  regime with a
single domain wall can be related
to a  class of driven critical systems which are
described by the stochastic transport equation with correlated quenched
currents \cite{BT-RG}.

In Section\ II we introduce the model and determine two disorder regimes.
In Section\ III numerical results of distributions and their  scaling
exponents are presented for varying  concentration $c$ and driving rates.
In Section\ IV we
discuss the analogy between low-disorder regime and stochastic transport
equation with quenched random currents,
and in Section\ V we give a short summary and discussion of the results.

\narrowtext
\begin{table}
\caption{Values of the scaling  exponents measured in various substances:
(MC1) amorphous iron, (MC2) ${\rm Ni_{95}Al_3Mn_2}$,
(MC3) ${\rm Fe_{78}B_{13}Si_{9}}$, all from Ref.1; (BG) VITROVAC
6025 X, Ref.6; (E4) ${\rm Fe_{30}Ni_{45}B_{15}Si_{10}}$, Ref.4;
(DS) ${\rm Fe_{64}Co_{21}B_{15}}$ under tensile stress, Ref.8;
 (TR)1.8\% ${\rm SiFe}$ for driving rate 0.15, Ref.5.
(D1) ${\rm Fe_{73}Co_{12}B_{15}}$ for driving rate 0.1, Ref. 7;
(DB) ${\rm Fe_{64}Co_{21}B_{15}}$ for driving rate 0.16, Ref.10. Except in
(MC1) and (MC2) which used wires, in the rest of experiments thin ribbon
or strip samples were used.}
\label{table1}
\begin{center}
\begin{tabular}{|c|c|c|c|c|c|c|c|c|c|}
   $E$& MC1&MC2&MC3& $BG$ & $E4$& $DS$& $TR$&D1&DB \\
\hline
$\tau _t $&  1.88&2.1&1.82& 2.22 & -& 1.5& 1.76& 1.72&1.66\\
$\tau _s$& 1.64&1.78& 1.74&1.77& 1.33 & 1.3&1.42&1.40&1.40 \\
$\tau _n$&1.44& 1.58&1.60& 1.56& -&-&-&-&-\\
$D_s$& 1.45&-&-& 1.51&-&-& 1.53&-&-\\
 $D_n$&-&-&-& 2.03&-&-&-&-&-\\
$\phi $&2&2&2& 1.6 & -& -& 1.75&-&-\\
\end{tabular}
\end{center}

\section{Model and Barkhausen avalanches}

We consider an Ising model on  the square lattice in
two-dimensions assuming that  local random fields $h_i$ are generated
 by  coarse-graining from an original disorder in the presence of
the external magnetic field $H$ \cite{BT}:
\begin{equation}
{\cal{H}} = -\sum_{<i,j>}J_{i,j}S_iS_j - \sum_i (h_i+H)S_i \ .
\label{Hamiltonian}
\end{equation}
$J_{i,j}=1$ is a constant interaction between nearst-neighbor spins
 $S_i=\pm 1$, and $J_{i,j}=0$ if at least one of the spins is zero.
A randomly distributed fraction $c$ of sites represents nonmagnetic ions
with $S_i=0$.  A Gaussian distribution of $h_i$ is assumed
with zero mean and width $f$. We assume
a small random field variance $f\le 0.3$ (see Ref.\cite{TN} for effects of
varying random fields), and vary concentration of nonmagnetic ions $c$.

Following Ref.\ \cite{TN}, we consider two disorder regions:
For low values of concentrations $c<c^\star(f)$ we
consider {\it extended domain wall} of flipped spins along
$<11>$ direction on the square lattice (linear size and the wall is $L$).
Periodic boundaries are applied in the direction of the wall, and  the
boundary opposite to the wall is left free. Here $c^\star (f)$ is determined
as the largest
disorder at which domain-wall depinning still occurs.
For high disorder $c>c^\star(f)$ the domain wall remains pinned for all
values of the driving field, and many domains of reversed spins are nucleated
inside the system. In this region of disorder periodic boundary conditions
in both directions can be used  equally well  as the boundary conditions
described above, leading  to the same avalanche statistics.
In Fig.\ 1 we show two examples of emerging cluster
structure of flipped spins in low-disorder and high-disorder regime.
 The cluster structure at low disorder reflects the  anisotropy
due to initial  conditions. The anisotropy
is   still important at high disorder, however, the  occurrence of
many domains  that block each other spatial extent becomes dominant feature
that determines the scaling properties of Barkhausen noise in this region.

We first apply infinitely slow driving (i.e., increments of the driving field
are adjusted to the minimum local field in the system). It should be
stressed that these driving conditions are useful for the purely theoretical
purposes, as opposed to finite driving rates that are applied in real
experiments.  An example of
the simulated BN signal for a half of the hysteresis loop is
given in Fig.\ 2. The corresponding time series of the values of
the field adjustments is  given in the inset to Fig.\ 2. It is clear
from Fig.\ 2 that
after a certain transient time the consecutive values of the weakest
local field in the system vary very little and, at the same time,
the main contribution to the avalanche statistics comes from exactly
that part of the hysteresis loop.

\section{Concentration and driving-rate dependence of scaling exponents}

In this Section we study distributions of avalanche sizes and durations
by keeping fixed weak random field $f\le $0.3 at each occupied site and vary
concentration of nonmagnetic ions $c$. We first apply infinitely slow driving
in order to differentiate effects of varying disorder from those of
varying driving rate. The concentration dependence for finite driving rate
in high-disorder region was studied in Ref.\ \cite{BT}.
At low concentrations $c$ the concept of a single domain wall persists,
although
a small fraction of spins may flip in the interior of the system near
vacancies. The domain wall moves in a series of slips
and eventually depinning occurs at critical driving field $H_c(c)$.
In the applied driving conditions the depinning may occur for concentrations
$c<c^\star (f)$, whereas for larger disorder the built-in domain wall
remains pinned for all values of the driving field. For $c>c^\star (f)$ the
domains of flipped spins are nucleated
inside the system and their walls  move at different instances of time.
In this way subsequent avalanches block each others extent, leading to
long-range correlations in space and to memory effects.
The two regions with different behavior are schematically shown in the phase
diagram in Fig.\ 3.

In the high disorder region HDR above $c^\star (f)$ we measure
the distribution of avalanche
durations, which is shown in Fig.\ 4 for varying $c$ and fixed $f=0.001$.
 Both the slope and
cutoff of the distributions depend on $c$.
In the inset are shown the extracted scaling exponents
for slopes of the distribution of durations ($\tau _t$), and sizes
($\tau _s$), which are found to increase with $c$.
By diluting the spin lattice clusters of connected spins become smaller
and more ramified, so that large avalanches are less probable
by fixed other conditions (i.e., fixed $f$ and driving rate). This leads
 to  the increase of  slopes of the avalanche distributions, as
shown in Fig.\ 4.

Similarly,  we find increase of the scaling exponents in the low
disorder region  (LDR) when $c$ is varied in the interval  $0<c<c^\star (f)$,
where a single domain wall exists. Here we have measured
the exponent of avalanche sizes $\tau _s(c)$, and derived $\tau _t(c)$
from the scaling relations that are discussed in detail in Ref.\ \cite{TN}.
The exponents are also shown in the inset to Fig.\ 4.
It is interesting to note that in this region of disorder the avalanches are
self-affine  (cf. Fig.\ 1).
We measure the anisotropy exponent $\zeta $, which is defined
as $<\ell _\bot>_{\ell _\|} \sim \ell _\|^\zeta $, where
$\ell _\|$ and  $\ell _\bot $ are avalanche extents in the direction
parallel and perpendicular to the wall, respectively. For instance, for
$c=0.05$ we find $\zeta =1.23 \pm 0.09$.

It is expected  that the depinning transition along the line $H_c(c)$
belongs to the same universality class as the depinning in $c=0$ limit
for low random-field type of disorder, which we studied in Ref.\ \cite{TN}.
When the disorder is tuned to a critical value along the line
$c^\star (f)$ the phase  transition occurs between
low-disorder  and high-disorder regime, which we expect to
be in the same universality class as the one at critical random-field
variance $f^\star$ for $c=0$ (see Ref.\ \cite{TN} for details).

By applying a finite driving rate, i.e., increasing the external field always
by a given constant amount $\Delta H$ independently of values of local
fields in the system, weak pinning centers become overdriven.
 Therefore avalanche sizes increase compared to the case of infinitely
slow driving
both due to coalescence of different clusters and by faster propagation of
 a single cluster due to reduced pinning.
As a result scaling exponents of avalanches decrease with increasing driving
rate (see Ref.\ \cite{BT2}). The effects of increased driving rate are
similar in both LDR and HDR, however, the boundary between these regions
in the phase diagram moves towards lower disorder (i.e.,
nucleation of new domains starts at lower disorder compared to the
case of zero driving rate and interaction between domain walls prevents
depinning transition). In the inset to Fig.\ 4 we have also shown
values of the scaling exponents in HDR for driving rate $\delta h
\equiv \Delta
H/H_{max}=0.01$, taken from Ref.\ \cite{BT}.
The concentration dependence of the exponents at finite driving rates
becomes  approximately linear (cf. inset to Fig.\ 4 and Ref.\ \cite{BT}).
Another important feature of the scaling behavior  by the finite
driving rate is that the cutoffs of the distribution curves increase
 and obey the finite-size scaling \cite{BT}.
In this way  a subcritical dynamic system for
$c^\star (f)<c<c_p(f)$, where $c_p(f)$ is the percolation threshold line,
 becomes critical {\it for a range of values of  driving rates} \cite{BT2}.
This is the reason for  ubiquity of self-similar Barkhausen noise
in highly disordered alloys, since the experimental setups
 always use finite frequency of the external field cycles.

Dependence of the scaling exponents of the finite driving rates has been
nicely demonstrated in Refs.\ \cite{TOR,Durin,Durin-stress,D+BG}
in 1.8\% Si-Fe and Fe-Co-B compounds.
It should be noted that both numerical  \cite{BT,BT2}
and experimental observations  \cite{TOR,Durin,Durin-stress,D+BG}
agree that for large driving rates  $\delta h$ the exponents decay
{\it linearly} with $\delta h$. However, for small driving rates
deviation from linear law occurs, making extrapolation to zero rate difficult.
The dependence of the measured exponents in the literature
on the concentration of nonmagnetic ions
can  be established  qualitatively. For instance, the duration exponent
 $\tau _t=1.56$ measured in ${\rm Fe_{73}Co_{12}B_{15}}$ (for
driving rate 0.3 in reduced units) in Ref.\ \cite{Durin} is lower
than $\tau _t=1.82$ measured
in diluted ${\rm Fe_{78}B_{13}Si_{9}}$  (for driving rate 1Hz)
in Ref.\ \cite{MC1},
in a qualitative agreement with our predictions (cf.inset to  Fig.\ 4).
However, a more quantitative comparison would be only possible if all
other conditions, i.e.,  driving rates and annealing of samples  were
kept under strict control.

\section{Universality class of driven disordered systems}

In the low-disorder regime  ($f<f^\star$, $c<c^\star (f)$)
apart from a weak $c$ dependence the critical behavior at infinitely
slow driving can be identified by the exponents $\tau _s=1.58$,
$\tau _t=1.89$ and  $\zeta =1.23$. These exponents appear to be close
to those obtained in the ricepile model  with stochastic critical slope rule
in 1+1 dimension \cite{ricepile}. It has been shown
\cite{PMB} that the same critical exponents occur in
a wide class of critical systems, that includes the models of
 ricepile, depinning and earthquakes.
The problem of  an interface depinned from quenched defects
in the {\it anisotropic} medium has been studied in  \cite{TKD}.
The exponents can be related to the directed percolation exponents
(see also \cite{TD}). It has been elucidated that due to anisotropy the
threshold driving force depends on the interface slope, and thus
may remain finite at the transition when the velocity of the interface
vanishes. The same effect is taken into account by the relevant nonlinear
term $\lambda (\partial _\|h)^2$ in the continuum equation
 of interface depinning \cite{KPZ}.
In addition, the   nonlinearity of the dynamics in critical systems
\cite{ricepile,PMB} leads to  self-tuning  to the dynamic  critical
state.

Recently it has been shown \cite{BT-RG} that a  self-tuning in the
class of
critical disordered  systems   can be described  by  self-consistent
building of correlations  in the transport equation with random
quenched currents
\begin{equation}
{{\partial h}\over{\partial t}}= \nu _{\Vert } \partial
^2_{\Vert }h +\nu _{\bot }\partial ^2_{\bot }h
- p(x)\partial _{\Vert }h +\eta \ .
\label{EOMO}
\end{equation}
Here  $p(x)$ is the  random  local  velocity of transport with zero mean
and  correlations given by
\begin{equation}
\langle p(x)\partial _{\Vert }h~ p(0) \partial _{\Vert }h
\rangle _d = \gamma
\left({{x_\bot^{1-\zeta }}\over{x_\|^{2-\delta }}}\right)
(\partial _{\Vert }h)^2,
\label{correlations}
\end{equation}
and the anisotropy exponent $\zeta $ is to be
determined self-consistently as the system approaches the fixed point.
The usual dynamic noise term $\eta $ is assumed to have annealed
correlations as $\langle \eta (x,t)\eta (x^\prime t^\prime)\rangle = 2D\delta
^{(d)}(x-x^\prime)\delta (t-t^\prime )$.
It has been found in Ref.\ \cite{BT-RG} that a stable fixed point exists
for finite strength of disorder $\gamma$ with the exponents that depend
continuously on the range of correlations along parallel direction,
parameter $\delta $. In particular, for long-range correlations at
$\delta =1/3$ the corresponding avalanche exponents are found $\tau _t=1.89$
and $\tau _s=1.56$, with  the emergent nonlinear term
${{x_\bot ^{1/9}}\over{x_\| ^{5/3}}}\left(\partial _\| h\right)^2$.

\section{Conclusions}

 Barkhausen  avalanches in disordered ferromagnets  obtained at
sufficiently slow (which respects time scale separation) increase of the
driving field reflect fractal structure  of the dynamic critical states.
The quantitative characterization of a given system  in terms of
sets of scaling exponents of the BN  avalanche can be expected in
several universality classes, which depend on: ($i$) type of domain walls;
($ii$) strength of pinning; and ($iii$) applied driving conditions.
The chemical composition, preparation, sample form and annealing  result
in certain type of domain walls, which are thus  characteristic
for a given sample. Whereas, the driving rate can be controlled in
the experimental setups. In ribbon geometry the
effects of demagnetizing fields are minimized with longitudinal
anisotropy \cite{LM}.

The scaling behavior studied in this work is
 expected to apply for the anisotropic samples with {\it extended}
domain walls, which are  pinned by nonmagnetic impurities.
 Such domain walls are found in ribbon samples with
strong longitudinal anisotropy
as for instance in fully annealed  ${\rm Fe_{78}B_{13}Si_{9}}$ ribbons
in which domain walls tend to orient along longitudinal anisotropy axis
\cite{LM,MC1}. Similar structure of domains was reported recently
in Refs.\ \cite{Durin-stress,D+BG}
for ${\rm Fe_{21}Co_{64}B_{15}}$ (see Fig.\ 1 in \cite{D+BG}).
Moreover, it has been suggested \cite{Durin-stress} that type of
domains in these samples can be controlled  by continuous variation of
the tensile stress.

Our results suggest (see also \cite{TN}) that self-similar
Barkhausen noise in these systems occurs  in a wide range of values of the
 disorder  and driving field, which are  marked as gray areas in the
 phase diagram in  Fig.\ 2, excluding only the area with finite domain-wall
velocity (upper left corner of the phase diagram). In the zero driving rate
we find two distinct universal behaviors for low disorder and high
disorder region, respectively, which compare well with the experimental
results in the above samples.
 For low disorder
self-similar BN is a consequence of the intermittent motion of a single
domain wall (or well separated, noninteracting domain walls),
 before eventually depinning of the wall occurs at critical
driving field.
We suggest that in the zero driving rate the BN in this region belongs to
the same universality class as the ricepile \cite{ricepile}
 and the anisotropic
interface motion in $1+1$ dimensions.
For high disorder region, which is  characterized by multidomain structure
in the plane (see for instance \cite{LM}), the criticality
of BN is {\it induced} by the finite driving rates, which represents
usually applied driving conditions  in the experiments of BN.
In principle, a {\it range of values of driving rates}
should exist for which a given sample exhibits self-similar Barkhausen noise.
Both numerical simulations \cite{BT2} and experimental observations
\cite{TOR,Durin,Durin-stress,D+BG} agree  that scaling
exponents decrease with increasing driving rate, with the
linear dependence at large driving rates. Discussion of the
finite driving rates in terms of the relevant fixed points of a transport
equation has not yet received full theoretical treatment.

\acknowledgments
This work was  supported by the Ministry
of Science and Technology of the Republic of Slovenia.


\narrowtext
\begin{figure}
\epsfxsize=82mm\epsffile[176 269 435 523]{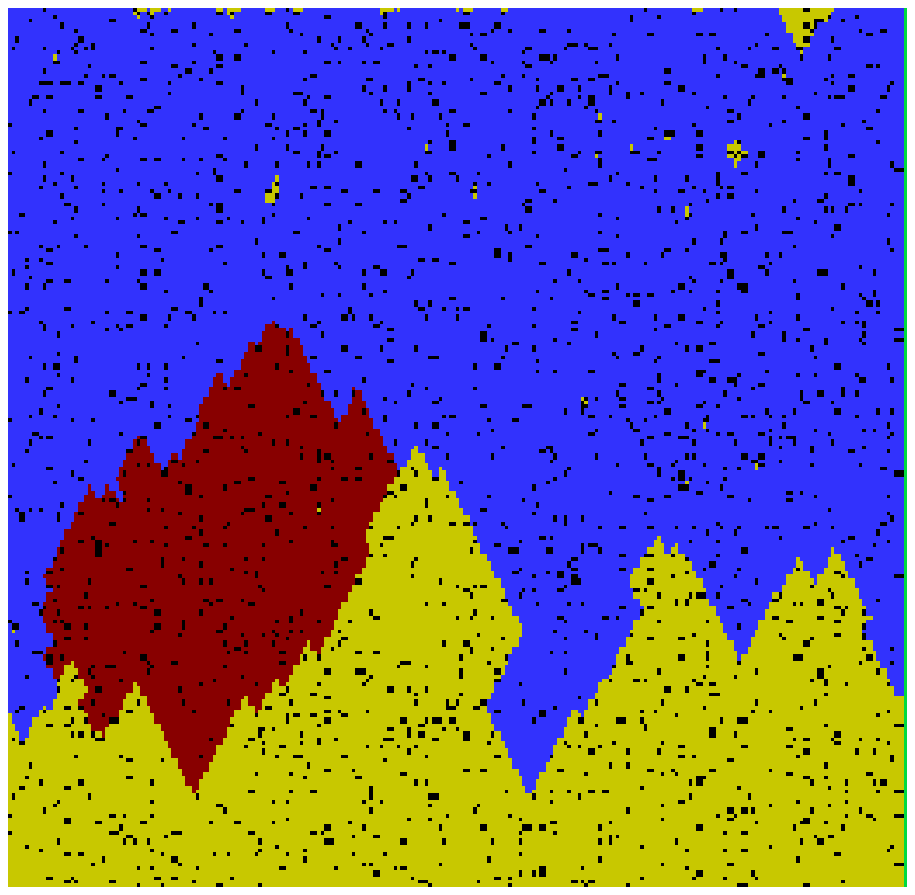}
\epsfxsize=82mm\epsffile[177 268 434 524]{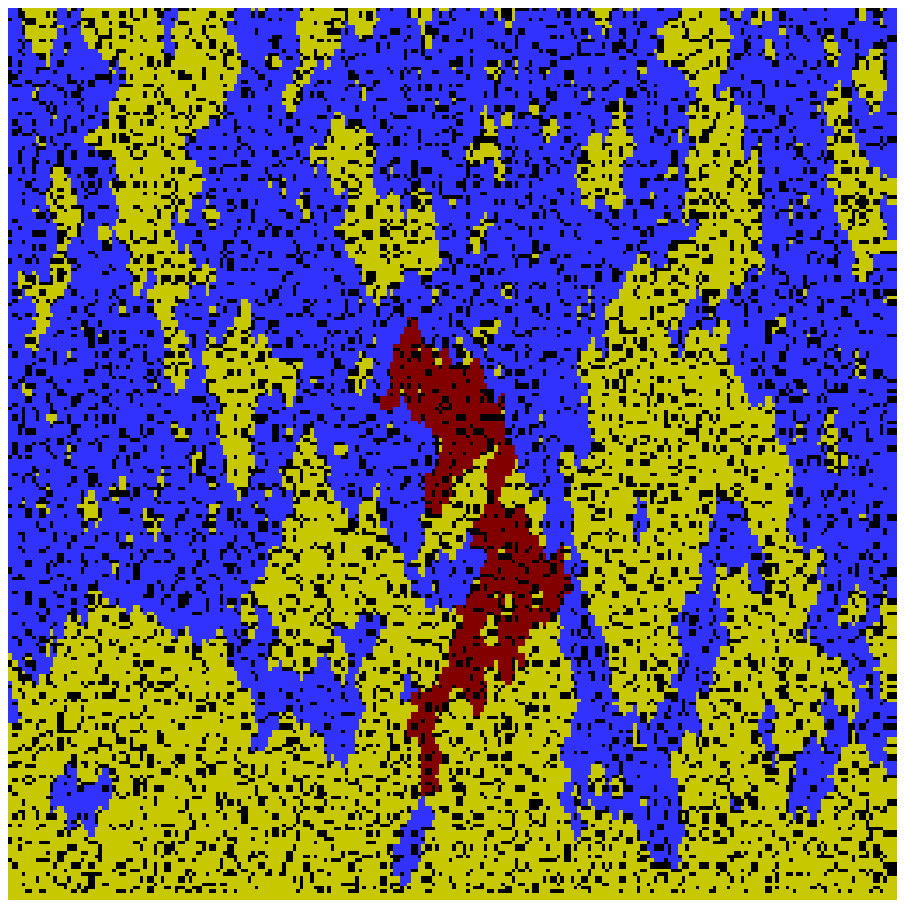}
\caption{Cluster of flipped (bright) and unflipped
(gray) spins, with most recent cluster (dark gray) for $L=200$ and $f$=0.3
in the presence of nonmagnetic defects (black points) of concentration $c$:
Bottom:  $c$=0.2; Top: $c$=0.05.}
\label{fig1}

\end{figure}

\begin{figure}
\epsfxsize=82mm\epsffile[38 68 522 581]{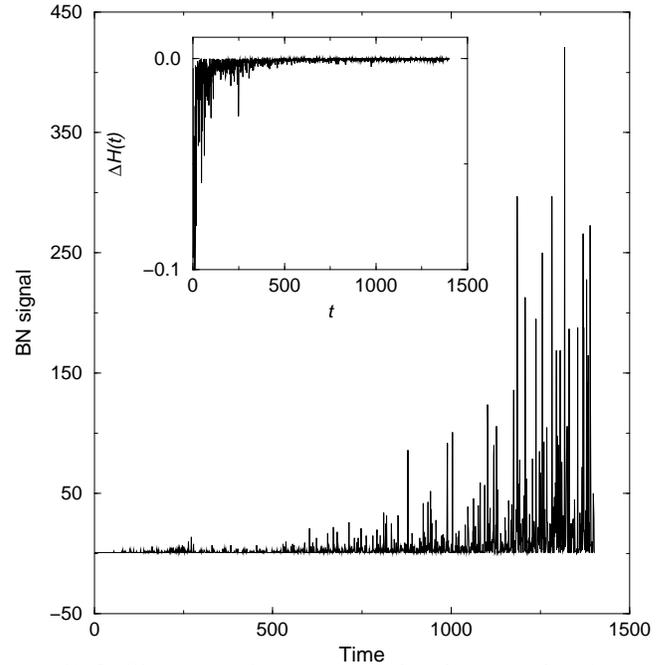}
\caption{Simulated Barkhausen noise signal vs time for one side of
the hysteresis loop. Lattice size is  $L=200$
and disorder values $f=0.3$ and $c=0.2$. Inset: Time series of the
increments of the external field $\Delta H(t)$ vs $t$ which  are
adjusted to the minimum local field in the system, corresponding to
the signal in  the main picture. Note that the main contribution to the
distribution of avalanches comes from the central part of the hysteresis
loop, which is reached after approximately 500 steps.}
\label{fig2}

\end{figure}
\begin{figure}
\epsfxsize=82mm\epsffile[47 68 522 501]{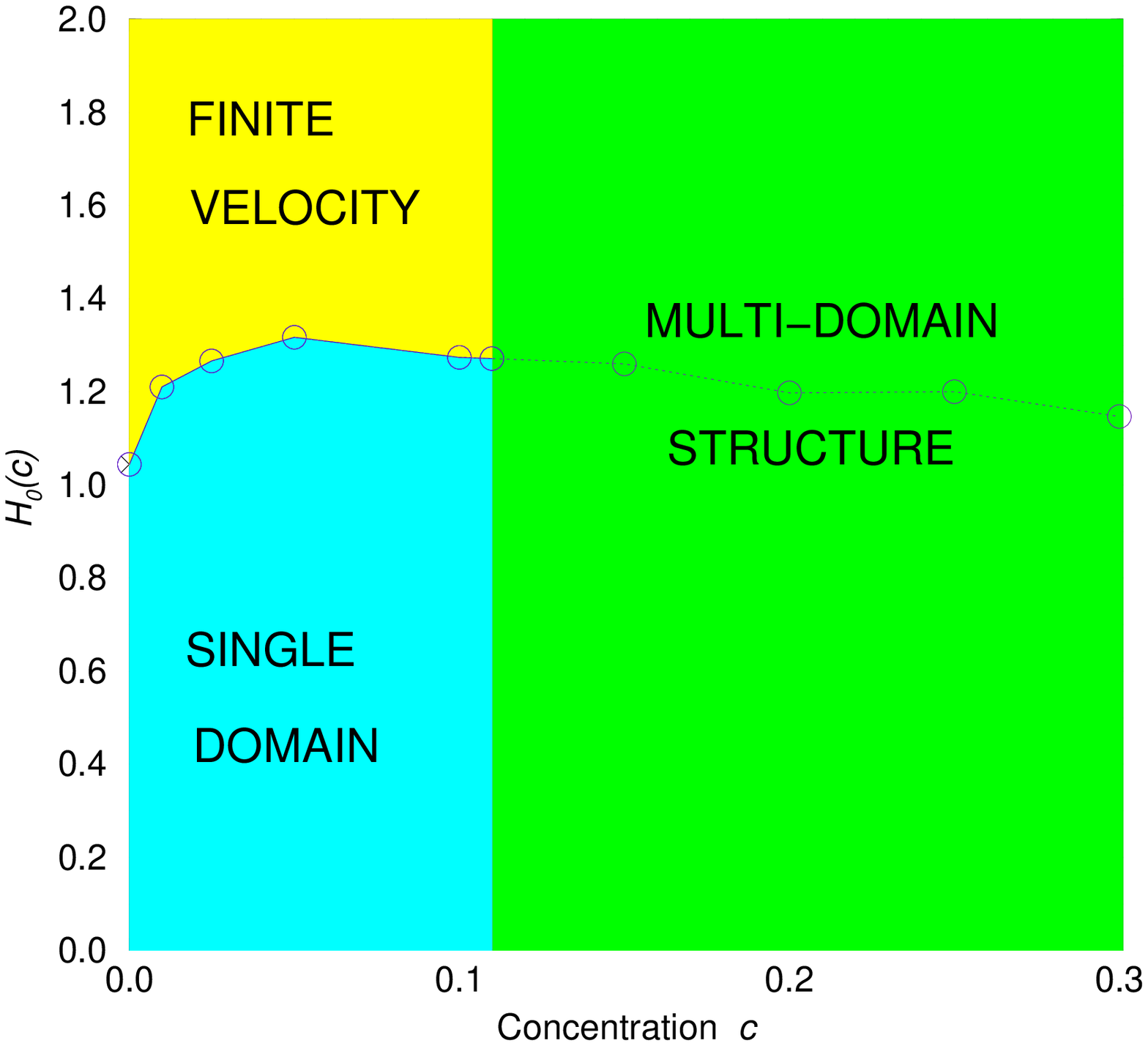}
\caption{Phase diagram: external field $H$ vs. concentration of
nonmagnetic ions $c$ for $f=0.3$ and infinitely slow driving.
Dashed line:  Coercive field $H_0(c)$ vs. $c$. Scale free Barkhausen
avalanche  (up to a finite cut-off) occur in the entire gray area,
corresponding either to single-domain or multi-domain structure.
 A single-domain depinning occurs along the boundary with finite
velocity region. For a finite driving rate boundary between two
scaling regions moves towards lower concentrations. }
\label{fig3}
\end{figure}


\begin{figure}
\epsfxsize=82mm\epsffile[47 68 522 581]{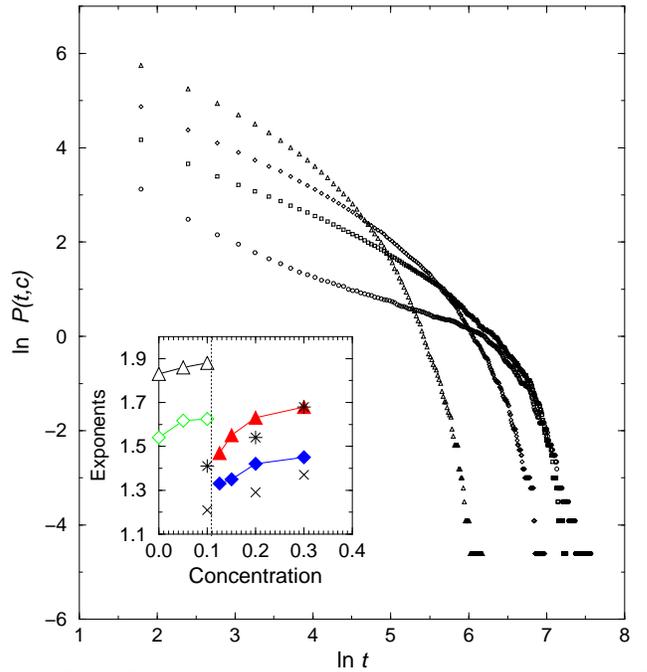}
\caption{Double logarithmic plot of the integrated
distribution of avalanche durations
$P(t,c)$ for  $f$=0.001 and various concentrations $c$=0.1,0.125, 0.15, and
 0.2,  obtained by
 infinitely slow driving for $L=200$. Inset: Critical exponents
($\triangle $) $\tau _t$, and ($\diamond $) $\tau _s$  vs. concentrations of
nonmagnetic ions $c$. Full symbols: high disorder; Empty symbols:
low disorder regime. Finite driving rate $\delta h =0.01$:
($\star $) $\tau _t$, and ($\times $) $\tau _s$ (from Ref.\ [14]). }
\label{fig4}

\end{figure}

\end{table}

\end{multicols}

\end{document}